\documentclass[aps,prb,epsfigm,twocolumn,showpacs]{revtex4-1}
\usepackage{graphicx}
\usepackage{amsmath}
\usepackage{amsfonts}

\begin{document}

\title{Theory of electrons, holes and excitons in GaAs polytype quantum dots}

\author{Juan I. Climente, Carlos Segarra, Fernando Rajadell, Josep Planelles}
 \affiliation{Departament de Qu\'{\i}mica F\'{\i}sica i Anal\'{\i}tica, 
 Universitat Jaume I, E-12080, Castell\'o, Spain}
 \email{josep.planelles@uji.es}
 \homepage{http://quimicaquantica.uji.es/}
\date{\today}

\begin{abstract}
Single and multi-band (Burt-Foreman) $k\!\cdot\!p$ Hamiltonians for GaAs crystal phase quantum dots
are developed and used to assess ongoing experimental activity on the role of such factors
as quantum confinement, spontaneous polarization, valence band mixing and exciton Coulomb 
interaction.
Spontaneous polarization is found to be a dominating term. Together with the control of
dot thickness [Vainorious \emph{et al.} Nano Lett. {\bf 15}, 2652 (2015)] it enables 
wide exciton wavelength and lifetime tunability.
Several new phenomena are predicted for small diameter dots 
[Loitsch \emph{et al.} Adv. Mater. {\bf 27}, 2195 (2015)], 
including non-heavy hole ground state, strong hole spin admixture 
and a type-II to type-I exciton transition, which can be used to
 improve the absorption strength and reduce the radiative lifetime 
of GaAs polytypes.
\end{abstract}


 \maketitle

\section{Introduction}

Semiconductor quantum dots (QDs) have been widely studied since the 1990s 
because of their appealing electronic and photonic properties.
However, standard fabrication methods involve a degree of dispersity 
which limits exact reproducibility within an ensemble of dots, from run-to-run and from lab-to-lab.\cite{TalapinJMC,BaumgardnerNS,FinleyPRB}
For example, in Stranski-Krastanov growth of InAs/GaAs QDs, the most widely employed
technique to produce optically active III-V QDs, random diffusion of substrate
material (GaAs) into the deposited material (InAs) leads to an ensemble of QDs with 
inhomogeneous composition, strain fields and shapes.\cite{FinleyPRB}
This translates into an inhomogeneous distribution of energy levels, which poses a challenge for 
the scalability of many technological applications demonstrated at a single-dot level.\cite{EconomouPRB,KimAPL}
%This is for instance the case 
%of optically controlled qubits, where the inhomogeneous distribution implies individually tuned lasers 
%to control each bit and restricts the ability to mediate entanglement via photonic cavity modes.\cite{EconomouPRB,KimAPL}

Crystal phase (polytype) QDs\cite{AkopianNL} are likely to mitigate this problem.
These structures exploit recent synthetic advances enabling control on the polytypical 
crystal structure of III-V nanowires, whereby one can grow alternating segments of wurtzite (WZ)
grown along the [0001] direction and zinc-blende (ZB) grown along [111].\cite{CaroffIEEE}
Because WZ and ZB phases have slightly different energy gaps at the $\Gamma$ point, 
band offsets are formed and carriers confined in one of the phases.\cite{SpirkoskaPRB}
One can then form QDs embedded in the wire, which turn out to have defect-free crystal structure, sharp interfaces,
negligible strain and tapering, well defined shape and homogeneous composition. 
Prospects have become especially promising with two studies published in the last months for GaAs polytype QDs.
On the one hand, Vainorious \emph{et al.} have reported exact control on the QD thickness, 
from bilayers to tens of nm.\cite{VainoriusNL}
On the other hand, Loitsch \emph{et al.} have reported control of the wire diameter from typical values
($\sim$100 nm) down to 7 nm.\cite{LoitschAM} 
Together, these studies pave the way towards full control of the QD confinement and, consequently,
of the energy structure.

Progress in the synthesis of GaAs polytype QDs, however, has not been paralleled by theoretical 
understanding of the ensuing electronic and optoelectronic properties. As a result, several 
open questions remain which need to be clarified in order to eventually attain predictive design.
To name a few: (i) the role of spontaneous polarization in WZ GaAs is not clear. The majority of 
experimental works simply neglect it\cite{SpirkoskaPRB,VainoriusNL,LoitschAM,CorfdirNL}, 
but recent theoretical\cite{BelabbesPRB} and experimental\cite{BauerAPL} 
studies point to a value of $P_{sp}=0.0023$ C/m$^2$, which Jahn \emph{et al.} deemed influential
at a single-particle level.\cite{JahnPRB} One then wonders if it is really important for Coulomb-bound
excitons. (ii) the role of valence band coupling is also poorly understood. It is generally assumed
that the hole ground state is a heavy hole (HH).\cite{VainoriusNL,LoitschAM,CorfdirNL} This is
consistent with polarization measurements of large diameter nanowires.\cite{SpirkoskaPRB2}
However, radial confinement enhances valence band mixing.\cite{EfrosPRB} Therefore, the validity should 
be tested at least in the small diameter regime enabled by the work of Loitsch \emph{et al}.\cite{LoitschAM}
(iii) the influence of electron-hole Coulomb interaction needs better assessment. 
Because WZ and ZB interfaces form a type-II band-alignment in GaAs\cite{SpirkoskaPRB}, 
previous simulations of optical transition energies in GaAs polytypes either tend to neglect it\cite{VainoriusNL,JahnPRB} 
or take it as a constant.\cite{LoitschAM} However, the band offsets are so small that both electron and hole
wave functions are expected to penetrate into each other's phase, leading to non-vanishing electron-hole overlap.\cite{CorfdirNL}
What is more, strong Coulomb interactions could eventually overcome the band offsets and change the excitons from 
type-II to type-I.  This is a possibility worth exploring. \\

In this work, we develop a $k\!\cdot\!p$ model to study carriers confined in polytype QDs, which takes into account 
all the factors described above: spontaneous polarization, electron-hole Coulomb interaction,
and valence band coupling of holes. The latter is included by building a 6-band Burt-Foreman Hamiltonian for 
ZB/WZ polytypes. Notice that the use of position dependent effective masses is convenient for GaAs due to 
the small band offsets.  We then study polytype QDs within the confinement ranges made possible by the
works of Vainorius\cite{VainoriusNL} and Loitsch\cite{LoitschAM}, and evaluate the influence of the
abovementioned factors on the energy structure of electrons, holes and excitons.
The results are discussed in view of the existing experiments.

\section{$k\!\cdot\!p$ Hamiltonians for WZ/ZB QDs} 

In order to model polytypes we need to obtain $k\!\cdot\!p$ Hamiltonians, spanned on the same Bloch functions, 
which are formally identical in both crystal structures, the differences showing up in the parameters only. 
In this section we derive such Hamiltonians for conduction band (CB) electrons, valence band (VB) holes and excitons.

%The local coordination of both WZ and ZB is the same: one kind of atom is tetrahedrally surrounded
%by the other kind of atom, the two structures being different mainly at the relative positions
%of the third neighbors and beyond. In fact, the layer stacking of the WZ crystal along the [0001]
%direction (ABAB...) corresponds to that of the ZB crystal along the [111] direction (ABCABC...),
%the difference holding in the dihedral conformation of the third layer, ``eclipsed'' and ``staggered'' for WZ
%and ZB, respectively.\cite{CaroffIEEE,YehPRB,FariaJAP}
%This similarity makes it possible to obtain approximate Hamiltonians which are formally identical in
%both phases, as shown below.

\subsection{Electrons}
\label{sec:ele}

Low-energy electrons in ZB GaAs belong to the $\Gamma_{6c}$ band, which is well separated from the valence
band and the rest of CBs. This justifies the widely spread use of single-band models 
in the literature. In WZ GaAs, however, $\Gamma_{8c}$ and $\Gamma_{7c}$ bands are close to each other
and some band mixing can be expected.\cite{DePRB,BechstedtJPCM} 
Lacking effective mass parameters describing such a coupling, we model WZ electrons with a single-band 
Hamiltonian of hybrid character: $\Gamma_{8c}$ masses but optically bright, like the
$\Gamma_{7c}$ band. This picture is consistent with the observations of different recent 
experiments\cite{CorfdirNL,SignorelloNC,GrahamPRB} and suffices to assess the role of the
physical factors we investigate.  The polytype Hamiltonian then reads:

\begin{equation}
\label{eqHe}
%H_{e}= -\frac{\hbar^2}{2} \sum_{i=\perp,z}
% \nabla_{i} \frac{1}{m^{*}_{i}} \nabla_{i} 
H_{e}= -\frac{\hbar^2}{2} \sum_{i=x,y,z}
 k_i \frac{1}{m^{*}_{i}} k_i
+ V_{c}^{cb}      
+ q V_{sp}.
\end{equation}

\noindent Here $m^*_i$ is the effective mass along the $i$ direction, 
which depends on the crystal phase, $k_i=-\nabla_i$, $V_{c}^{cb}$ is the 3D confinement 
potential arising from the conduction band-offset potential between ZB and WZ phases,
$q$ is the electron charge and $V_{sp}$ is the electrostatic potential due to the spontaneous 
polarization $P_{sp}$. 

The calculation of strain in polytype QDs deserves a short discussion. 
The initial strain in a heterostructure is given by the lattice mismatch.
For a QD of a given material buried in a matrix of a different material with the 
same crystalline structure, it is zero in the matrix and $\epsilon^0_{ii} = \frac{a_i^{(m)}-a_i^{(QD)}}{a_i^{(m)}}$ 
in the QD, where $a_i^j$ is the lattice constant in the direction $i$ for the medium $j$.\cite{RajadellJAP} 
However, this expression cannot be employed in polytypes because QD and matrix have different crystalline structure. 
In our case, since we deal with ZB(111)/WZ(0001) interfaces, we may reason as follows. 
The ZB unit cell contains 9 anions and 9 cations while the WZ unit cell has the same basis but different height and only 
6 pairs of ions (see e.g. Fig.1 in Ref.~\onlinecite{FariaJAP}). Then, three WZ unit cells contain the same number of ions 
as two ZB unit cells. If the ZB and WZ materials are the same (GaAs in our case) and under the assumption that the 
lattices are ideal, the basis surface of both unit cells is the same, and so is the height of three WZ unit cells vs. two ZB ones. 
Therefore, the strain is ideally zero. This is consistent with theoretical calculations\cite{LahnemannPRB} and 
experimental findings\cite{VainoriusNL,LoitschAM,LahnemannPRB} pointing at negligible strain, as real lattice constants
show but small deviations from ideal ones. One can then safely disregard it.

Since the strain is weak, so is the piezoelectric potential and its influence on the energy spectrum. 
By contrast, the spontaneous polarization potential $V_{sp}$ can have a significant influence.
There is no spontaneous polarization in the ZB phase for symmetry reasons, but it is present in WZ,
where $P_{sp}$ originates from the ``eclipsed'' dihedral conformation of layers $N$ and $N+2$, 
yielding a non-ideal tetrahedral coordination and associated electric dipoles.
Current estimates for GaAs are of $P_{sp}\approx 0.0023\,Cm^{-2}$,\cite{BelabbesPRB,BauerAPL} 
about one order of magnitude weaker than in nitride materials.
Since the change in $P_{sp}$ is large at the ZB/WZ interface, it gives rise to an abrupt change in the built-in 
electric field, from zero in ZB up to an approximate constant value $F$ in WZ given by  $F = P_{sp}/\varepsilon$, 
with $\varepsilon$ the dielectric constant, and back again to zero in ZB (see e.g. Fig.~4 in Ref.~\onlinecite{BauerAPL}). 
Then, the QD acts like a capacitor, with effective negative and positive charges accumulating
at the ZB/WZ and WZ/ZB interfaces, and an almost linear potential in between (see e.g. CB profile in insets of Fig.~\ref{fig2}). 
%to a minimum negative value at the ZB/WZ interface where effective negative charges arise. 
%Then an abrupt almost linear increase of the potential up tu reaching the same value, but now positive, 
%at the next WZ/ZB interface and a further decreasing in the ZB phase, appraching zero again  
%(see e.g. Fig. 2 in \cite{park2}). 

\subsection{Holes}
\label{sec:hole}

\subsubsection{Multi-band Hamiltonian}

To study the effect of VB mixing we use a multi-band $k\!\cdot\!p$ Hamiltonian.
In order to compare [111]-grown ZB and [0001]-grown WZ structures systematically,
we write the six-band Hamiltonian for both phases using the basis functions with
lower symmetry, which are those used for WZ crystals:\cite{FariaJAP,ParkJAP2}
%We take advantadge of the similarity between [0001]-grown WZ and [111]-grown ZB lattices 
%by using the so-called cubic approximation.\cite{Birpikus_book,SuzukiPRB,ChuangPRB}
%This allows us to describe the both WZ and ZB phases using a WZ Hamiltonian, with a particular
%set of parameters in each phase.
% 

\begin{equation}
\begin{array}{ll}
\label{eq1}
|u_1\rangle =-\frac{1}{\sqrt{2}} |(X + i \, Y) \uparrow\rangle & |u_4\rangle =\frac{1}{\sqrt{2}} |(X -i \, Y) \downarrow\rangle \cr
\cr
|u_2\rangle =\frac{1}{\sqrt{2}} |(X - i \, Y) \uparrow\rangle & |u_5\rangle =-\frac{1}{\sqrt{2}} |(X +i \, Y) \downarrow\rangle \cr
\cr
|u_3\rangle = |Z \uparrow\rangle & |u_6\rangle =|Z \downarrow\rangle \cr .
\end{array}         
\end{equation}

\noindent For [0001] WZ, the resulting Hamiltonian in this basis reads:\cite{ChuangPRB}
\begin{equation}
\label{eq2}
H_{6B} = 
 \left[  \begin{matrix} F &-K^* & -H^* & 0 & 0 & 0  \cr
         -K & G & H & 0 & 0 & \sqrt{2} \Delta_3  \cr  
		 -H & H^* & \lambda & 0 & \sqrt{2} \Delta_3 & 0  \cr
		 0 & 0 & 0 & F & -K & H\cr
		 0 & 0 & \sqrt{2} \Delta_3 & -K^* & G & -H^*  \cr
		 0 & \sqrt{2} \Delta_3 & 0 & H^* & -H & \lambda 
		 \end{matrix}\right],
\end{equation}

\noindent where
 
\begin{eqnarray}
\label{eq3}
F &=& \Delta_1+\Delta_2 +\lambda+\theta \nonumber \\
G &=& \Delta_1-\Delta_2 +\lambda+\theta \nonumber \\
\lambda &=& \frac{\hbar^2}{2 m_e}[A_1 k_z^2+ A_2 k_{\perp}^2]  \nonumber \\
\theta &=& \frac{\hbar^2}{2 m_e} [A_3 k_z^2+ A_4  k_{\perp}^2] \nonumber \\
K &=& \frac{\hbar^2}{2 m_e} \, A_5 k_+^2 + \Delta K\\
H &=& \frac{\hbar^2}{2 m_e} \, A_6 k_+ k_z + \Delta H            \nonumber.
\end{eqnarray}

\noindent Here $m_e$ is the free electron mass, $A_i$ effective mass parameters,
$k_\perp=k_x^2+k_y^2$, $k_\pm=k_x\pm i k_y$, $\Delta_1$ is the crystal field splitting, 
$\Delta_2$ and $\Delta_3$ spin-orbit matrix elements,
and $\Delta K=\Delta H=0$.

For $[111]$ ZB, the $[001]$ ZB Hamiltonian --spanned on the basis of Eq.~(\ref{eq1})-- 
is first rotated $45^\circ$ along the $z$ axis, and then $54.7^\circ$ along the new $y'$ axis.
The resulting $z'$ axis points along the $[111]$ direction, while $x'$ and $y'$ do so
along the [$11\bar{2}$] and [$\bar{1}10$] directions. The Hamiltonian obtained is formally 
identical to Eq.~(\ref{eq2}), but now:
 
\begin{eqnarray}
\label{eq4}
\Delta K &=& 2 \sqrt{2} \, \frac{\hbar^2}{2 m_e} \, A_z k_- k_z     \nonumber  \\
\Delta H &=& \frac{\hbar^2}{2 m_e} \, A_z k_-^2                  
\end{eqnarray}

\noindent Additionally, the following relations emerge, which reduce the number of independent 
mass parameters to three, as expected for ZB:

\begin{equation}
\begin{array}{l}
\label{eq5}
\Delta_1=0 \\
\Delta_2=\Delta_3 = \Delta/3 \\
A_1=-\gamma_1-4 \gamma_3 \\
A_2=-\gamma_1+2 \gamma_3 \\
A_3=6 \gamma_3 \\
A_4=-3 \gamma_3 \\
A_5=- \gamma_2-2 \gamma_3 \\
A_6=-\sqrt{2} \, (2 \gamma_2+\gamma_3) \\
A_z=\gamma_2-\gamma_3.
\end{array}         
\end{equation}

\noindent where $\gamma_1,\,\gamma_2,\,\gamma_3$ are the Luttinger mass parameters.
 
It is worth noting that $H_{6B}$ --with ZB parameters, Eqs.~(\ref{eq4}) and (\ref{eq5})--,
 actually shows all diagonal elements overstabilized by an amount $\Delta/3$.
 This is due to the term $\frac{1}{3} \Delta (\sigma \cdot \mathbf J)$ 
in the sum of invariants defining $H_{6B}$,\cite{Birpikus_book} which is needed to yield the 
Hamiltonian extradiagonal elements $H_{26}$, $H_{35}$, $H_{53}$ and $H_{62}$. 
We recover the zero origin at the top of the HH band by subtracting $\Delta/3$ to all diagonal elements of
$H_{6B}$ in the ZB region. \\

\noindent The above considerations prompt us to obtain a Hamiltonian which is valid for both ZB and WZ regions,
and hence open the possibility of dealing with polytypes. 
 Since the parameters in the two phases are different, we should employ a variable mass Hamiltonian.\cite{mireles,veprek_prb,veprek_opt,schubert} According to Burt-Foreman,\cite{foreman_prb} this requires adding some extra coefficients:

\begin{equation}
\label{eq6}
H_{6B}^{BF} = \left[  \begin{matrix} F-\rho & \kappa & \xi^* & 0 & 0 & 0 \cr
         \kappa^* & G+\rho & -\xi & 0 & 0 & \sqrt{2}\Delta_3  \cr  
		 \eta & -\eta^* & \lambda & 0 & \sqrt{2}\Delta_3 & 0 \cr
		 0 & 0 & 0 & F+\rho & \kappa^* & -\xi\cr
		 0 & 0 & \sqrt{2}\Delta_3 & \kappa & G-\rho & \xi^* \cr
		 0 & \sqrt{2}\Delta_3 & 0 & -\eta^* & \eta & \lambda
		 \end{matrix}\right]
\end{equation}

\noindent where

\begin{eqnarray}
\label{param4}
F &=& \Delta_1+\Delta_2 +\lambda+\theta \\
G &=& \Delta_1-\Delta_2 +\lambda+\theta \nonumber \\
\lambda &=&\frac{\hbar^2}{2 m_e} \; \left[k_z A_1 k_z+ k_x A_2 k_x+ k_y A_2 k_y \right] \nonumber \\
\theta &=& \frac{\hbar^2}{2 m_e} \; \left[k_z A_3 k_z+ k_x A_4 k_x+ k_y A_4 k_y \right] \nonumber \\
\kappa &=& \frac{\hbar^2}{2 m_e} \; \left[-k_x A_5 k_x+ k_y A_5 k_y+ i\; (k_x A_5 k_y+k_y A_5 k_x) \right]+\Delta \kappa \nonumber \\
\eta &=& \frac{\hbar^2}{2 m_e} \; \left[-k_z A_6^{(+)} k_+ - k_+  A_6^{(-)} k_z\right]+\Delta \eta \nonumber \\
\xi &=& \frac{\hbar^2}{2 m_e} \; \left[-k_z A_6^{(-)} k_+ - k_+  A_6^{(+)} k_z\right]+\Delta \xi \nonumber \\
\rho &=& \frac{\hbar^2}{2 m_e} \; \left[i\; k_y \, (A_5^{(+)}-A_5^{(-)}) k_x - i\; k_x \, (A_5^{(+)}-A_5^{(-)}) k_y\right] \nonumber \\
\Delta \xi &=& \frac{\hbar^2}{2 m_e} \; \left[-(k_x-i\; k_y) A_z (k_x-i\; k_y)\right] \nonumber \\
\Delta \eta &=&  \Delta \xi \nonumber \\
\Delta \kappa &=& -2 \sqrt{2} \, \frac{\hbar^2}{2 m_e}\, \left[ (k_x+ i \; k_y) \, A_z^{(+)} \, k_z+ k_z \, A_z^{(-)} \,(k_x+i\; k_y)\right], \nonumber
\end{eqnarray}

\noindent with $A_5= A_5^{(+)}+A_5^{(-)}$,  $A_6= A_6^{(+)}+A_6^{(-)}$ and $A_z= A_z^{(+)}+A_z^{(-)}$. 
In the WZ region, $A_z^{(+)}=A_z^{(-)}=0$. In the ZB region, Eqs.~(\ref{eq5}) still hold.
\\

\noindent The practical hindrance for the use of this Hamiltonian, especially for studying polytypes, 
is the lack of available $A_i^{(\pm)}$ coefficients. 
At this regard, Veprek et al.\cite{veprek_prb} analyzed the spurious solution problem affecting the $k\!\cdot\!p$ envelope function method, 
which is related to the lack of ellipticity that the different sets of parameters confer to the Hamiltonian.
They concluded by recommending the use of a complete asymmetric operator ordering  ($A_i^{(+)} =A_i$, $A_i^{(-)}=0$) 
for several ZB and WZ materials.\cite{veprek_prb, veprek_opt, zhou} We have cheked that this also applies to GaAs.

We are now in a condition to write the complete Hamiltonian for holes in polytype QDs:

\begin{equation}
\label{eqH}
H_h^{6B} = H_{6B}^{BF} + (V_c^{vb} -q V_{sp} - \frac{\Delta}{3} Y_{ZB}) {\mathbb I}_{6\times 6}.
\end{equation}

\noindent
$V_c^{vb}$ and $V_{sp}$ are the confining and spontaneous polarization potentials,
respectively, which we obtain as described for electrons. $Y_{ZB}$ is a heaviside function,
$Y_{ZB}=0$ in the WZ phase and $Y_{ZB}=1$ in the ZB one.

\subsubsection{Single-band Hamiltonian} 

The use of single-band models for the hole ground state in GaAs polytypes is justified under certain conditions.
In ZB, the degeneracy between heavy and light hole bands can be lifted by quantum confinement.
In WZ, as can be seen in Eq.~(\ref{eq2}), the uppermost band ($F$) is split from the others
($G$, $\lambda$) by the spin-orbit ($\Delta_2$) and crystal field ($\Delta_1$) splittings, so degeneracy
is lifted even at the $\Gamma$ point.
Thus, in order to get the single-band Hamiltonian, we decouple the diagonal elements 
corresponding to the $F$ (heavy hole) band from the rest of the matrix in Eq.~(\ref{eqH}).
This yields:

%\begin{equation}
\begin{multline}
%H_{11} = \frac{\hbar^2}{2 m_e} \, \left( k_z \frac{1}{m_z} k_z + k_{\perp} \frac{1}{m_\perp} k_{\perp} \right) + V_c^{vb}. 
\label{eqH1B}
%H_{h}= \Delta_1 + \Delta_2 + \sum_{i=\perp,z}
%\frac{\hbar^2}{2} \nabla_{i} \frac{1}{m^{*}_{i}} \nabla_{i} 
H_{h}= \Delta_1 + \Delta_2 + \sum_{i=x,y,z}
\frac{\hbar^2}{2} k_{i} \frac{1}{m^{*}_{i}} k_{i} 
+ V_{c}^{vb}      
- q V_{sp} - \frac{\Delta}{3} Y_{ZB},
\end{multline}
%\end{equation}

\noindent where the parameters $\Delta_1$, $\Delta_2$ and $m^*_i$ take different values in each
crystal phase. In particular, for WZ, $m_z=1/(A_1+A_3)$ and $m_\perp=1/(A_2+A_4)$.
For ZB, $m_z=-1/(\gamma_1-2\gamma_3)$ and $m_\perp=-1/(\gamma_1+\gamma_3)$,
$\Delta_1=0$ and $\Delta_2=\Delta/3$.\cite{improve_mass}

\subsection{Excitons}

We calculate neutral excitons using single-band Hamiltonians for both electron and hole:

\begin{equation}
\label{eqX}
H_X = H_e + H_h + V_{eh},
\end{equation}

\noindent where $V_{eh}$ is the electron-hole Coulomb interaction, which is
obtained by integrating the Poisson equation in a dielectrically inhomogeneous environment.

\subsection{System and Parameters}

We take ZB GaAs material parameters from Ref.~\onlinecite{VurgaftmanJAP}. 
As for WZ GaAs, we take electron effective masses from Ref.~\onlinecite{DePRB}, VB ones
from Ref.~\onlinecite{CheiwchanchamnangijPRB} and the spontaneous polarization $P_{sp}= 0.0023\,C m^{-2}$ from Ref.~\onlinecite{BelabbesPRB}. 
Lacking more precise information for WZ, we use a dielectric constant $\varepsilon=13.18$
for both phases.\cite{AdachiJAP}
%and the remaining parameters from Ref.~\onlinecite{BarettinIEEE}. 
 
\begin{figure}[h]
\begin{center}
\includegraphics[width=0.45\textwidth]{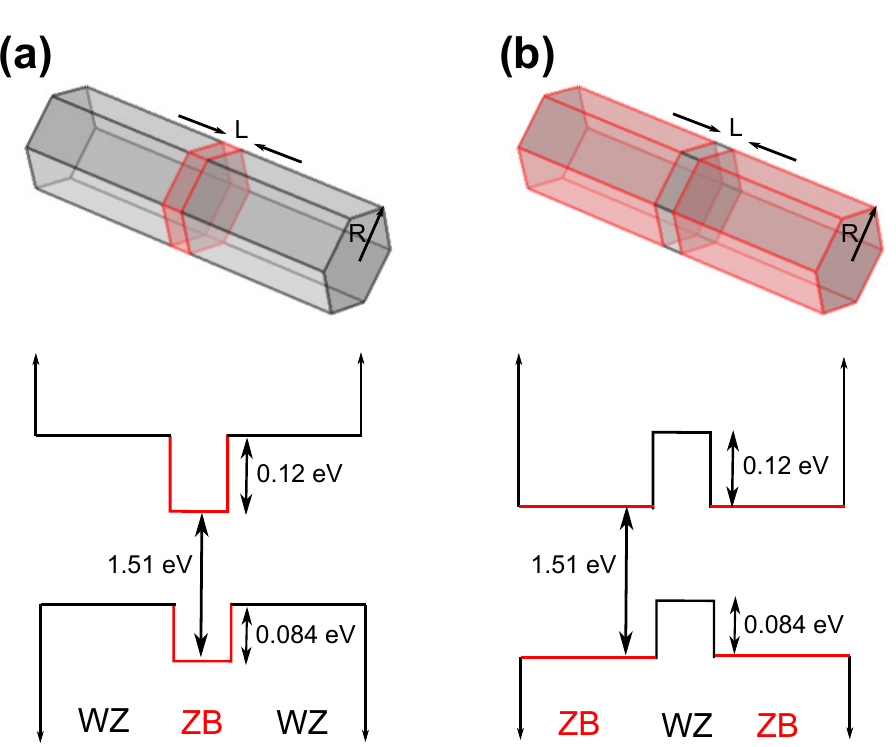}
\caption{(Color online). (a) Sketch of ZB QD embedded in WZ wire and corresponding band-offset profile.
(b) Same but for WZ QD embedded in ZB wire.}
\label{fig1}
\end{center}
\end{figure}

To define $V_c^{cb}$ and $V_c^{vb}$, we consider either ZB QDs embedded in WZ nanowires, 
see Fig.~\ref{fig1}(a), or WZ QDs embedded in ZB nanowires, see Fig.~\ref{fig1}(b). 
The corresponding band offset values, represented in the figure, 
are taken from Ref.~\onlinecite{SpirkoskaPRB}. 
The nanowires are assumed to be surrounded by an insulating material with 
$V_c^{cb}=-V_c^{vb}=5$ eV and $\varepsilon=4$. 

We use Comsol 4.2 to solve numerically the Hamiltonians described above.
$V_{sp}$ is obtained by calculating the polarization charge density  
$\rho=-\nabla \cdot P_{sp}$ and solving the Poisson equation,
$ \nabla \cdot [\varepsilon(\mathbf{r}) \; \nabla V]=\rho$,
where $\varepsilon(\mathbf{r})$ is the position-dependent dielectric constant.
Next, Hamiltonians $H_e$, $H_h^{6B}$ and $H_h$ are integrated using finite elements.
As for $H_X$, converged interacting electron and hole states are obtained by
using an iterative Schr\"odinger-Poisson scheme.
Note that the electrostatic potential obtained by integrating the Poisson equation 
accounts for the polarization originated by the inhomogeneous dielectric medium.
On the other hand, we disregard the self-polarization potential because it barely has
influence along the nanowire axis and, in particular, on the heterointerface,
as the dielectric mismatch between the two GaAs phases is negligible. 
Only in wires with small radius, the mismatch with an insulating environment 
could bring about additional radial confinement, although similar in both phases.

\section{Results}

\subsection{Electrons}

\begin{figure}[h]
\begin{center}
\includegraphics[width=0.45\textwidth]{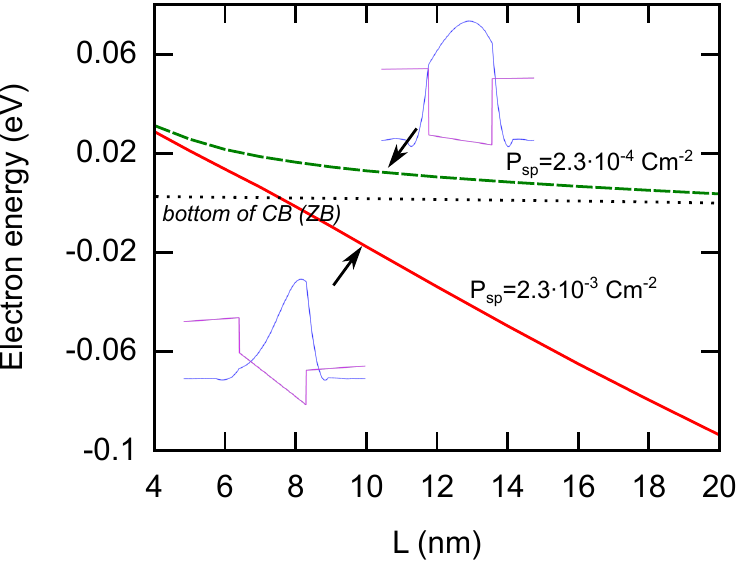}
\caption{(Color online). Energy of the electron ground state in a ZB QD with realistic (solid line) and weakened (dashed line) 
spontaneous polarization, as a function of the dot thickness. Note the strong influence. 
The insets show the wave functions and band profiles for $L=10$ nm. }
\label{fig2}
\end{center}
\end{figure}

We start by investigating the ground state of a single electron in a ZB QD, % embedded in a WZ wire,
like that in Fig.~\ref{fig1}(a).
The QD is hexagonal, with a typical radius of the circumscribed circle, $R=50$ nm, and variable thickness $L$. 
The results are plotted in Fig.~\ref{fig2}, where we compare calculations with the expected 
spontaneous polarization of GaAs, $P_{sp}= 2.3\cdot10^{-3}\,C m^{-2}$ (solid line), and calculations with 
an artificially weakened polarization, $P_{sp}= 2.3\cdot10^{-4}\,C m^{-2}$ (dashed line).
It is clear from the figure that, except for thin dots ($L < 5$ nm), 
$P_{sp}$ plays a critical role in determining the electron energy. 
Under full polarization, the energy shows a linear dependence with $L$, 
determined by the capacitor-like built-in electric field. %$F \approx P_{sp}/\varepsilon$.
By contrast, under weakened polarization, the linear regime is preceded by a quadratic one (up to $L\lesssim 10$ nm), 
which is determined by quantum confinement.
The magnitude of the energy stabilization is also very different. 
In fact, for $P_{sp}=2.3\cdot10^{-3}\,C m^{-2}$ and large $L$, the electric field 
leads to energies well below the CB bottom of bulk ZB. 

These results are qualitatively similar to those obtained by Jahn and co-workers using 
a simpler 1D model with zinc-blende masses,\cite{JahnPRB} and confirm that the spontaneous 
polarization in GaAs polytypes cannot be neglected, at least at a single particle level.
As we shall see below, in Section \ref{s:exc}, the same is true for excitons in spite of 
the electron-hole attraction.

The insets in Fig.~\ref{fig2} show the electron wave function and band profile for the
full and weakened polarization values. Notice that $P_{sp}$ pushes the electron towards
the ZB/WZ interface and induces substantial spreading into the WZ phase.\\

\begin{figure}[h]
\begin{center}
\includegraphics[width=0.45\textwidth]{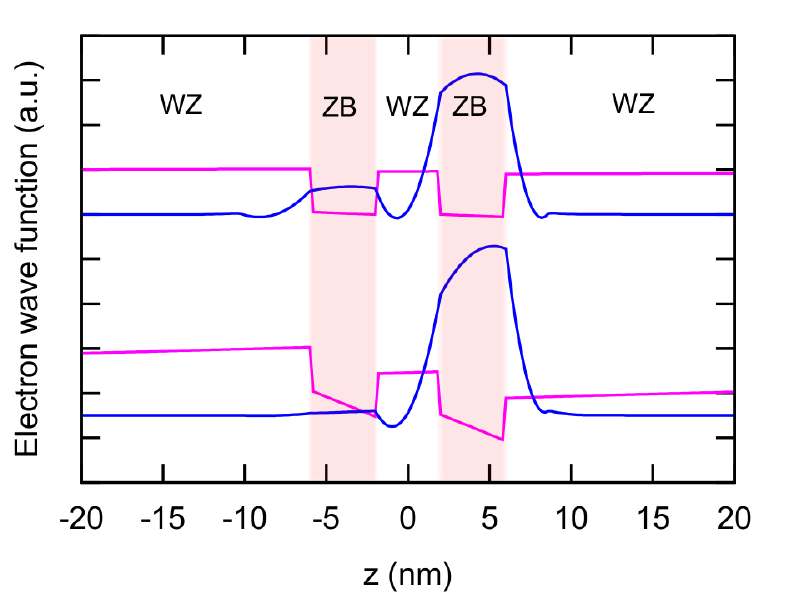}
%\caption{Electron wave function in (a) a ZB-WZ-ZB molecule, and (b) a ZB-WZ one.
%Note that the second structure partially restores the symmetry lifted by $P_{sp}$.
\caption{(Color online). Electron wave function and band profile in a ZB-WZ-ZB molecule with weakened and full $P_{sp}$.
Both ZB QDs and the WZ barrier have thickness $L=4$ nm.}
\label{fig3}
\end{center}
\end{figure}

The precise control of the thickness in GaAs polytype QD, recently achieved by 
Vainorius \emph{et al},\cite{VainoriusNL} suggests such structures could be used to build perfectly 
symmetric pairs of QDs. In principle, this could enable the formation of 
QD molecules with homonuclear character, unlike in self-assembled InAs/GaAs structures where the 
inherent structural asymmetries can only be overcome with external fields.\cite{BrackerAPL}
Symmetric molecules can be of interest for applications like optical qubits\cite{BayerSCI} 
or the development of superlattices with maximized coherent tunneling for solar cell devices.\cite{TomicAPL}
However, the results of Fig.~\ref{fig2} suggest that the strong influence of $P_{sp}$ 
can introduce significant asymmetries in the band profile of symmetric molecules. 
This is confirmed in Fig.~\ref{fig3}, where one can see that for two identical ZB QDs
separated by a thin WZ barrier, the electron wave function localizes mostly in one of the dots.
This effect is already noticeable if the system has weakend $P_{sp}$ (upper plot), and it becomes
dramatic for full $P_{sp}$ (lower plot), when tunneling is almost nearly suppressed.

%In order to partially restore the symmetry, we propose an alternative configuration,
%by which a ZB dot is placed next to a WZ one of the same thickness, as plotted in 
%Fig.~\ref{fig3}(b). The symmetric placement of two WZ/ZB interfaces around the 
%central ZB/WZ one, gives rise to a symmetric potential $V_{sp}$. 
%Since GaAs band offsets are small, this grants substantial electron tunneling.

\subsection{Holes}

%The influence of QD confinement on the hole energy is similar to that of electrons 
%reported above.
The energetics of holes in WZ QDs is qualitatively similar to that of electrons in ZB dots. 
In this section, we focus on the role of VB mixing instead.
In particular, we assess the validity of the usual assumption that the ground state 
has a well defined, single-band, HH character.\cite{VainoriusNL,LoitschAM,CorfdirNL}
 
The eigenfunctions of Hamiltonian $H_h^{6B}$ are six-component spinors of the form:
$\Psi_h^{6B} = \sum_{i=1}^6 f_i(\mathbf{r}) |u_i\rangle$, where $f_i(\mathbf{r})$ 
is the envelope function associated with the $|u_i\rangle$ Bloch function.
The weight of an individual component is computed as $|f_i|^2$. 
As can be seen in Eq.~(\ref{eq1}), HH character corresponds to Bloch
functions $|u_1\rangle$ (spin up) and $|u_4\rangle$ (spin down), i.e. the $F$ band
of Hamiltonian $H_{6B}^{BF}$.
To disentangle spin up and down components, 
a small Zeeman-like term is included in Eq.~(\ref{eqH}), $\Delta_z = B \mu_B g {\mathbb J}_z$,
where $B=1$ T is the longitudinal magnetic field, $\mu_B$ the Bohr magneton,
$g=4/3$ the hole $g$-factor and ${\mathbb J}_z$ the angular momentum 
$z$-component diagonal matrix (with elements $\pm 3/2$, $\pm 1/2$). % esto supone un splitting a 1 T ~0.24 meV (casi)

\begin{figure*}[t]
\begin{center}
\includegraphics[width=0.75\textwidth]{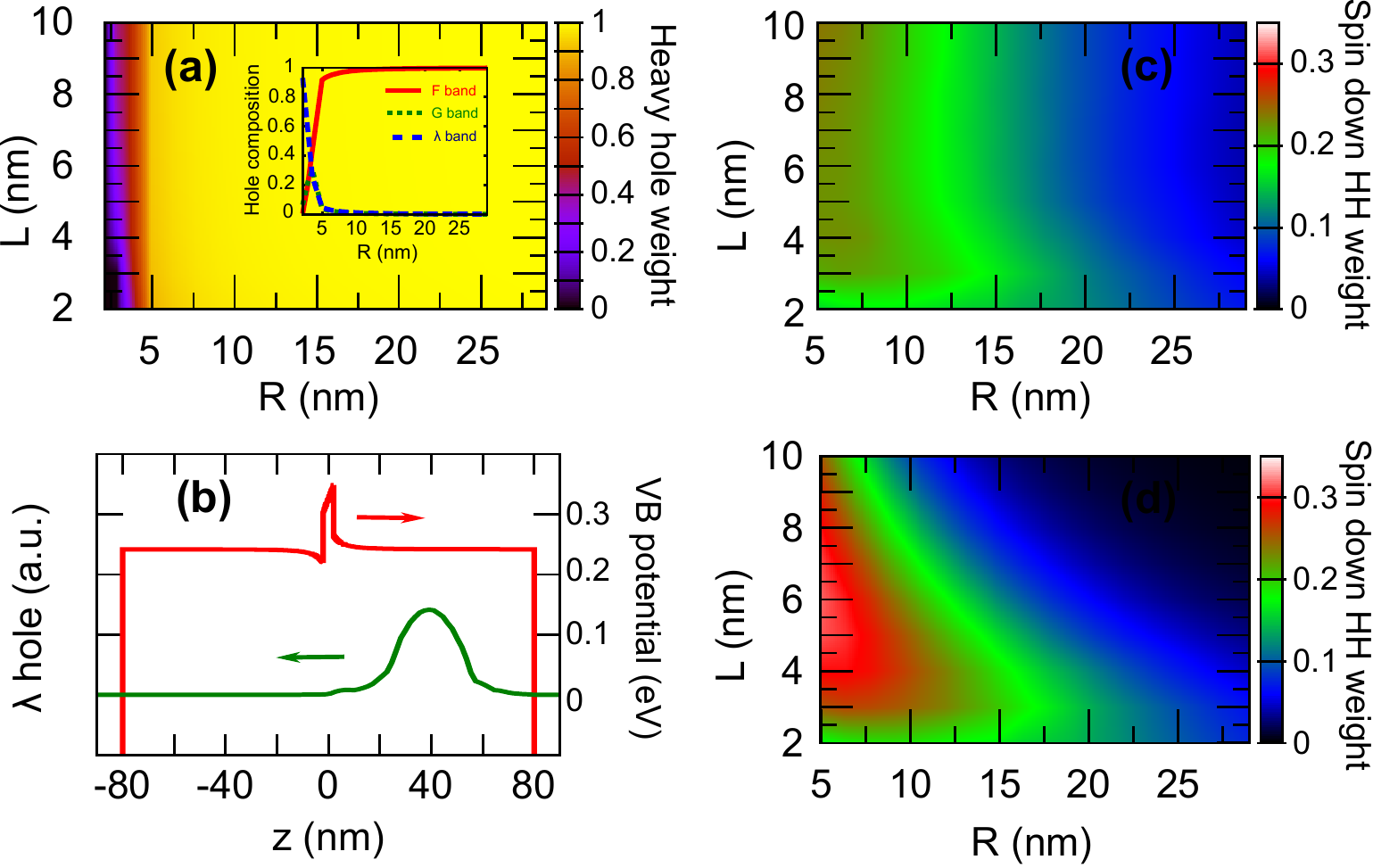}
\caption{(Color online). (a): Weight of the HH ($F$ band) component in the hole ground state, $(|f_1|^2+|f_4|^2)/\sum |f_i|^2$, 
as a function of dot radius and thickness in a WZ QD.
The inset shows the weight of each subband for $L=4$ nm. The ground state changes from $F$ band (HH) to $\lambda$ band for radii under 5 nm.
(b) Confining potential (red) and $\lambda$ hole envelope wave function (green) along the $z$ axis for a QD with $L=4$ nm and $R=2.5$ nm.
The $\lambda$ hole is not confined owing to its small mass.
(c) and (d): Weight of spin down HH component, $|f_4|^2/\sum |f_i|^2$,
 with and without $P_{sp}$, respectively.}
\label{fig4}
\end{center}
\end{figure*}

We first estimate the total HH character from $(|f_1|^2$+$|f_4|^2$). 
Fig.~\ref{fig4}(a) shows the normalized HH weight for the ground state of WZ QDs.
The QDs structure is that of Fig.~\ref{fig1}(b), with the radius $R$ and thickness 
$L$ varying within a parameter space enabled by state-of-the-art fabrication,
which includes the regime of quantum confinement in the radial direction.\cite{VainoriusNL,LoitschAM}
One can see that the ground state has almost exclusive HH character
except for very thin radii, $R < 5$ nm. 
In this region, the ground state rapidly switches from 
mainly $F$ band (HH) to mainly $\lambda$ band character, as shown in 
Fig.~\ref{fig4}(a) inset.
The origin of the ground state change can be understood as follows.
In the bulk limit, the $F$ band of wurtzite is stabilized with respect to 
$G$ and $\lambda$ bands by the crystal field and spin-orbit splittings.
However, the radial mass of $F$-band holes in WZ is $m_\perp^F=1/(A_2+A_4)=-0.13$,
much lighter than that of $\lambda$-band holes, $m_\perp^\lambda=1/A_2=-0.617$.
Therefore, with increasing radial confinement, the latter become more stable.
It is worth noting that $\lambda$ holes are very light in the [0001] direction,
$m_z^\lambda=1/A_1=-0.05$. As a consequence, their kinetic energy exceeds the small
band offset of the WZ/ZB interface, and the wave function tends to localize
outside the QD, as shown in Fig.~\ref{fig4}(b). 

The change of ground state character we report here should be observable in the
narrowest wires synthesized by Loitsch \emph{et al}.\cite{LoitschAM} 
Since the symmetry of the $\lambda$ band Bloch functions 
($|u_3\rangle$ and $|u_6\rangle$ in Eq.~(\ref{eq1})) is different
from that of HHs, it should be seen in experiments as a change
in the polarization of interband optical transitions.

Next we analyze the HH spin admixture by representing the ratio between the weight of spin up and down HH,
$|f_1|^2/|f_4|^2$. This is done for QDs in the presence and absence of $P_{sp}$, 
Figs.~\ref{fig4}(c) and (d) respectively.
A remarkable observation is that moderate radial
or vertical confinement leads to significant spin mixing between the Zeeman split levels. 
This is because the off-diagonal elements of Hamiltonian (\ref{eq6}) scale with $k_\perp$
and $k_z$. Note that, in the presence of $P_{sp}$, the spin mixing takes place even for large $L$,
 because the vertical electrostatic confinement does not vanish with the dot thickness.
 Interestingly, the confinement leads to admixture between spin up and down $F$-band holes
but coupling with $G$ and $\lambda$ bands remains negligible (recall Fig.~\ref{fig4}(a) inset). 
% ($|f_1|^2+|f_4|^2 \approx 1$).

\begin{figure}[h]
\begin{center}
\includegraphics[width=0.45\textwidth]{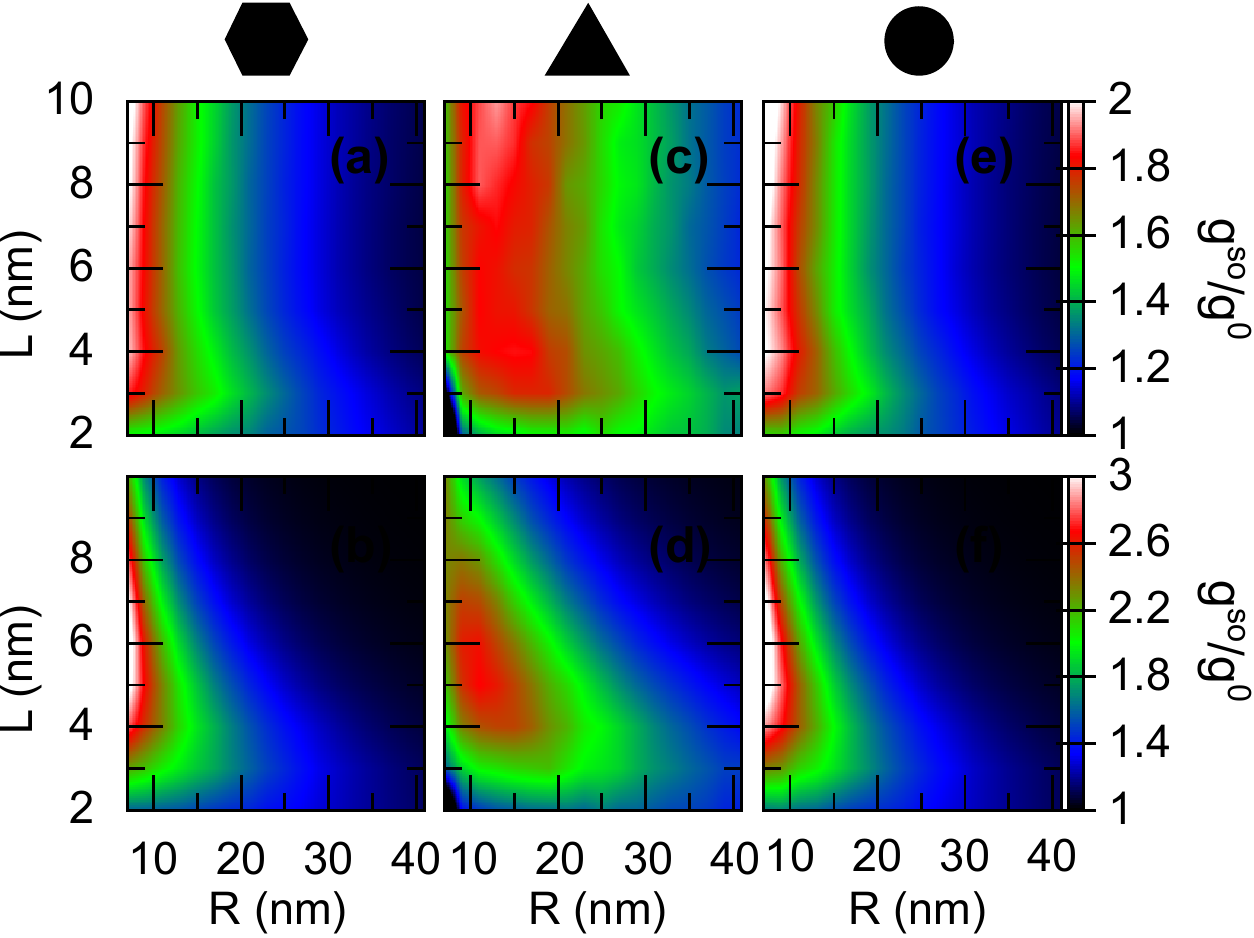}
\caption{(Color online). Ratio of hole $g$-factors calculated with and without VB mixing, 
$g^{so}/g^0$. (a) and (b): hexagonal QDs in the presence and absence of $P_{sp}$, 
respectively. (c) and (d): same but for triangular QDs. (e) and (f) same but for
cylindrical QDs. For the sake of comparison, the nominal radius refers to the 
circumscribed circle of the triangle. For hexagons and cylinders the actual radius 
is scaled so as to preserve the same area as the triangle.}
\label{fig5}
\end{center}
\end{figure}

The spin mixing of HHs has influence on the magnetic properties of WZ/ZB QDs.
For example, the values of the effective $g$-factors are often determined 
as $g=(E_+(B)-E_-(B))/\mu_B B$, where $E_\pm(B)$ is the energy of opposite spin projections under 
a magnetic field $B$. We have calculated the $g$ factor at $B=1$ T switching on and off
the spin-orbit term $\Delta$ in Hamiltonian (\ref{eq6}). This leads to $g$ factors with
enabled ($g^{so}$) and suppressed ($g^0$) spin mixing. 
In Fig.~\ref{fig5} we plot the ratio $g^{so}/g^{0}$  
in two instances: QDs with (a) and without (b) spontaneous polarization.
One can see that for strong confinement, band mixing can enhance $g$ factor values
up to a factor of 2-3. 
Unlike in other QD systems, where the confinement symmetry plays a critical role
in determining the spin admixture strength,\cite{PlanellesPRB} 
the mixing in polytypes is robust against symmetry changes,  
as similar numbers are obtained if one replaces hexagonal 
wires by triangular\cite{ZouSMALL} (Fig.~\ref{fig5}(c) and (d)) or cylindrical 
(Fig.~\ref{fig5}(e) and (f)) ones.

Recent experiments with GaAs polytype QDs revealed strong dispersion of the measured 
excitonic $g$-factors depending on the confinement strength.\cite{CorfdirNL}
As Fig.~\ref{fig5} shows, due to the VB mixing, the hole $g$-factor value can 
fluctuate substantially for different QD dimensions. 
This may partially explain the experimental observation.

One concludes from this section that the single-band HH description of the ground
state is valid except for small diameter wires, when a $\lambda$-band ground state is formed. 
However, in the presence of magnetic fields, one should be aware that VB mixing can 
strongly couple spin up and down HH states. 
We stress that such a spin mixing is mediated by excited (light-hole like) 
$G$ and $\lambda$ states, although they barely couple to the ground state themselves. 
In fact, as we show in Appendix \ref{sec:effH}, the mixing cannot be described with effective 
two-band Hamiltonians.  It is an intrinsic many-band coupling effect. 

\subsection{Excitons}
\label{s:exc}

In what follows we investigate the influence of confinement and spontaneous polarization
on the properties of the ground state exciton. In order to compare with available experiments, 
we restrict to radii $R\geq 5$ nm, where the single-band HH description is valid. 
The exciton state is thus calculated with Eq.~(\ref{eqX}), which fully accounts for electron-hole Coulomb interaction.

\begin{figure}[h]
\begin{center}
\includegraphics[width=0.48\textwidth]{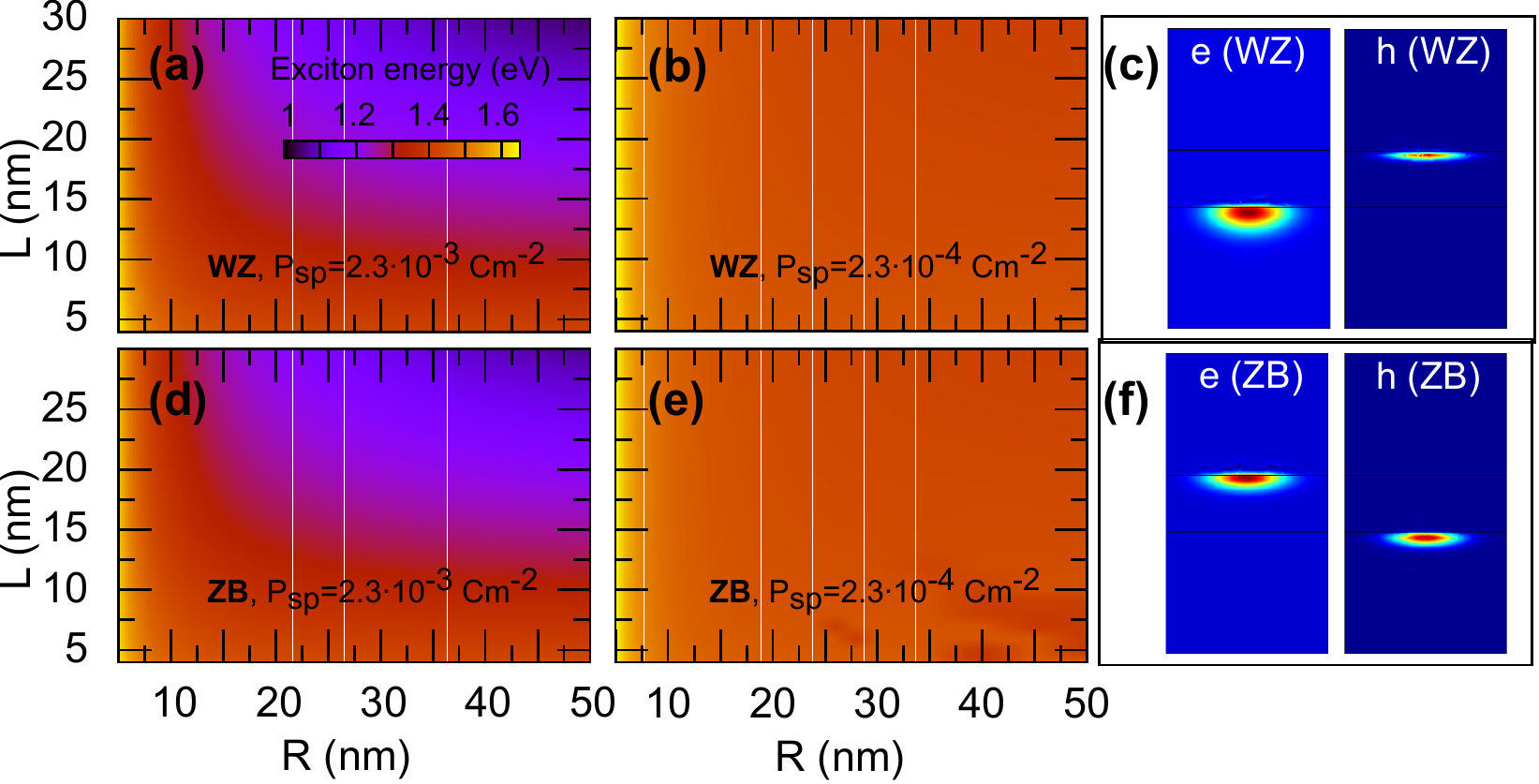}
\caption{(Color online). Exciton energy as a function of radius and thickness in
WZ QDs with full $P_{sp}$ (a) or weakened $P_{sp}$ (b).
(c) Excitonic electron and hole wave functions for a WZ QD
with $(R,L)=(50,30)$ nm and full $P_{sp}$.
(d)-(f): same but for ZB QDs.}
\label{fig6}
\end{center}
\end{figure}

Figure \ref{fig6} shows the exciton energy in WZ QDs embedded in ZB wires, panels (a) and (b), 
and ZB QDs embedded in WZ wires, panels (d) and (e). 
The left column corresponds to full spontaneous polarization, $P_{sp}=2.3\cdot10^{-3}\,C m^{-2}$, and the
right one to artificially weakened polarization, $P_{sp}=2.3\cdot10^{-4}\,C m^{-2}$.\cite{why}
One can see that for the realistic value of $P_{sp}$, the exciton energy has a very strong dependence 
on both the QD radius and thickness. 
The wavelength tunability is actually remarkable, as the energy can be tuned by over 700 meV, 
well above and below the bulk band gap ($1.51$ eV), from $1.65$ eV (visible) to $1.0$ eV (near infrared). 
For weak $P_{sp}$, instead, the tunability is reduced. Radial quantum confinement still plays a role,
enabling exciton emission up to $1.65$ eV for the narrowest wires.
By contrast, the influence of the dot thickness is largely suppressed, 
as in the single-particle case we saw in Fig.~\ref{fig2}.
Consequently, the lowerbound exciton emission is only $1.41$ eV, roughly the indirect band gap 
between the bottom of the ZB CB and the top of the WZ VB, see Fig.~\ref{fig1},
which is the smallest possible energy allowed by quantum confinement alone.
%In fact, in the absence of $P_{sp}$, quantum confinement alone 
%This is because vertical quantum confinement, which is now dominant, cannot reduce the gap below that
%has a weaker influence on the energy 
%than the electrostatic potential $V_{sp}$, as we saw in Fig.~\ref{fig2}.

 It is worth noting that for large thicknesses, the electron-hole overlap decreases.
This effect is especially pronounced in the presence of full $P_{sp}$, as shown in 
Figs.~\ref{fig6}(c) and (f), where it can be seen that the exciton electron and hole
wave functions localize at opposite interfaces.
For this reason, the exciton becomes gradually dark and optical experiments may not
be able to observe low energy states.

\begin{figure}[h]
\begin{center}
\includegraphics[width=0.45\textwidth]{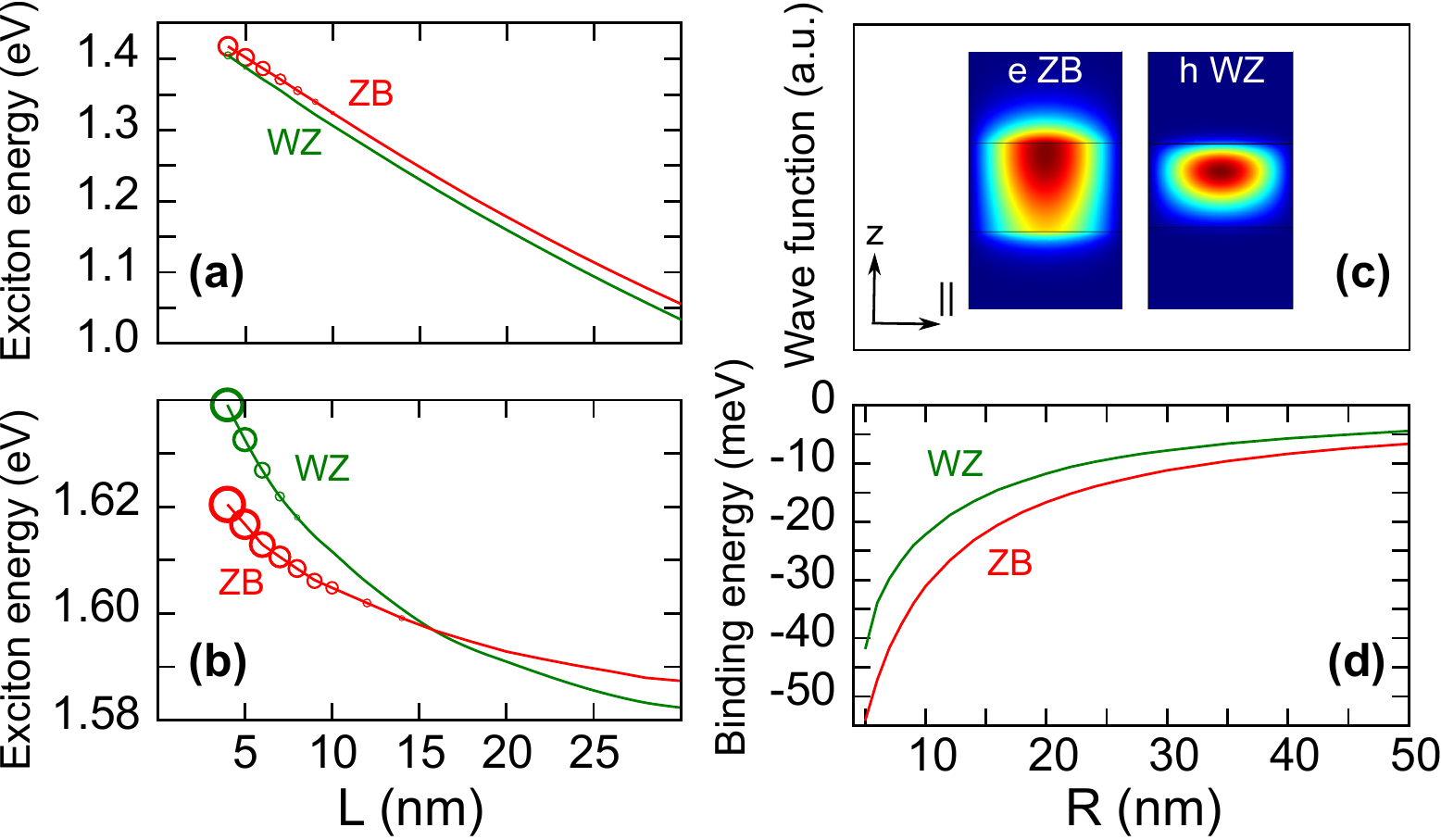}
\caption{(Color online). (a) Exciton energy vs thickness in WZ (green) and ZB (red) QDs with $R=50$ nm. 
(b) Same but with $R=5$ nm.  Note the qualitative change in behavior between the two radii.
The size of the circles is proportional to the electron-hole envelope function overlap. 
(c) Wave function of exciton's electron in a ZB QD (left) and hole in a WZ QD (right), 
both with $(R,L)=(5,5)$ nm. (d) Exciton Coulomb energy vs QD radius, with $L=4$ nm.}
\label{fig7}
\end{center}
\end{figure}

To better compare the optical properties of WZ QDs and ZB QDs, in Fig.~\ref{fig7} we plot
the exciton energy as a function of the dot thickness. Two representative cases are considered.
In Fig.~\ref{fig7}(a) we study typical QDs, with large radius, $R=50$ nm and full $P_{sp}$.
In this case, the thickness dependence is linear for both WZ and ZB, owing to to the large built-in 
electric field coming from $P_{sp}$, as already noted for electrons in Fig.~\ref{fig2}.
It follows that spontaneous polarization prevails over Coulomb interactions in GaAs polytypes.
This validates similar theoretical predictions obtained for ZB QDs at a single-particle level,\cite{JahnPRB}
which here we extend to WZ QDs.
Besides, we find that excitons in WZ QDs (green line) have lower energy than those in ZB QDs (red line), 
regardless of $L$.
This is due to the smaller kinetic energy of the confined carrier, as for holes in WZ $m_z=0.89$, % tmp *** update w Cheiwchancham params?
while for electrons in ZB $m_z=0.067$. For the same reason, holes leak out of the QD to a lesser extent
than electrons. Consequently, the electron-hole overlap --proportional to the size of dots 
in Fig.~\ref{fig7}--  is also weaker for WZ QDs.

Interestingly, the behavior described above changes drastically when one switches to QDs with strong 
radial confinement. This can be seen in Fig.~\ref{fig7}(b), which corresponds to QDs with $R=5$ nm.
First, the thickness dependence becomes quadratic in spite of $P_{sp}$. This is because the radial
confinement provides enough energy for the carriers to escape from the electrostatic potential wells.
The resulting wave functions are no longer localized near the WZ/ZB interface, but rather all over the QD, 
see Fig.~\ref{fig7}(c). Hence, they become sensitive to the quantum confinement in the
growth direction.  Second, ZB becomes the lowest emitting structure for thin dots ($L<15$ nm).
The origin of this inversion can be inferred from Fig.~\ref{fig7}(c) as well. 
The electron in the ZB QD can compensate for the strong radial confinement by penetrating into the WZ region, 
but the hole in the WZ QD cannot. This is again due to the relative masses $m_z$ of the confined carrier. 
 %which is smaller for the electron than for the hole.
Third, the electron-hole overlap is enhanced with respect to that of large diameter QDs, for both WZ and ZB 
(compare the size of the circles in Fig.~\ref{fig7}(a) and (b)).
This is also connected with the confined carrier delocalizing all over the QD and penetrating 
into the wire crystal phase, which reduces the separation from the outer carrier.
In other words, radial confinement induces a gradual transition from the usual type-II
behavior of GaAs polytype QDs to a type-I one.
One then concludes that reverse reaction growth\cite{LoitschAM} can be used for
 structural designs which improve the absorption strength and reduce the radiative lifetime 
of GaAs polytypes, as previously proposed for other (spontaneous polarization free) materials.\cite{ZhangNL,self}

We note that the type-II to type-I transition is partially stimulated by Coulomb interaction. 
As shown in Fig.~\ref{fig7}(d), the exciton binding energy scales up with radial confinement. 
For small radii, it reaches several tens of meV, the same order of magnitude as the ZB/WZ band offsets.
 As a result, it helps bring electron and hole closer together in the growth direction.

Before closing this section, we briefly discuss the relation of the results in 
Fig.~\ref{fig7} with those of available experiments.
Vainorius and co-workers recently measured the pholuminescence of WZ and ZB 
QDs with variable thickness and large radii, $R\approx 45-60$ nm.\cite{VainoriusNL}
They found that WZ emission is redshifted with respect to ZB one by a few tens of meV, 
in agreement with our Fig.~\ref{fig7}(a). However, in their experiments the emission 
energy was less sensitive to $L$. From $L=5$ nm to $30$ nm, the exciton emission
in ZB QDs redshifted by about $60$ meV, one order of magnitude less than we predict.
They noted that a similar redshift could be obtained in theory if one considers
quantum confinement only. We have confirmed this with our model, by considering 
excitons under weakened polarization (not shown).\cite{extra}
The question then arises of whether the experimental system had, for some reason, 
suppressed spontaneous polarization at the ZB/WZ GaAs interface.
Having uncapped wires, migration of surface charges might play a role in this sense.
Indeed, the authors indicated that surface effects could be responsible for fluctuations 
of tens of meV among QDs of the same thickness but different diameter. 
New experiments with capped GaAs/AlGaAs wires should study the linear or quadratic
dependence on $L$ to confirm the role of spontaneous polarization.
On the other hand, we expect the electron-hole overlap to decrease with $L$, see Fig.~\ref{fig7}(a). 
For thick QDs, it might be that the exciton ground state is dark and experiments are measuring 
emission from excited states, whose overlap is greater.\cite{extra2}

As for the radius dependence, the experiments of Loitsch {\emph et al.} reported exciton
emission up to $1.61$ eV for WZ QDs with $R\approx5$ nm and random thickness, which is 
blueshifted by $100$ meV with respect to bulk GaAs. %This is close to our estimates for 
%thick dots in Fig.~\ref{fig7}(b), $L>15$ nm. 
According to our estimates in Fig.~\ref{fig7}(b), for thin dots one could reach even
stronger blueshifts. In addition, for $R=5$ nm the experiments measured excitonic 
lifetimes under 1 ns, much shorter than the several ns of large diameter QDs.\cite{SpirkoskaPRB}
This can be taken as a first experimental evidence of the type-II to type-I transition we predict
with increasing radial confinement. New experiments systematically comparing ZB and WZ dots with 
small diameter would be useful to confirm the other new phenomena we predict in this regime, namely the 
change of the VB forming the hole ground state under $R=5$ nm ($\lambda$ band), and the fact that ZB 
QDs emit at lower energies than WZ ones for short $L$.

\section{Conclusion}

We have developed a $k\!\cdot\!p$ model to investigate WZ/ZB polytype QDs, 
including 3D confinement, spontaneous polarization effects, 
VB coupling through a Burt-Foreman six-band Hamiltonian,
and electron-hole Coulomb interaction for excitons.
When applied to GaAs QDs, we find a number of relevant observations:
\renewcommand{\theenumi}{(\roman{enumi})}%
\begin{enumerate}
  \item contrary to what is often assumed, spontaneous polarization in GaAs is 
	not negligible; it should dominate the electronic structure for QDs with 
	thickness above $\sim 5$ nm.
  \item the hole ground state has nearly pure HH character except
	for the narrowest wires, $R<5$ nm, when it switches to $\lambda$ band.
  \item when subject to a magnetic field, the HH ground states experiences 
	strong spin mixing, mediated by excited valence bands.
  \item strong radial confinement brings about a transition from indirect
	(type-II) to direct (type-I) excitons, and partially masks spontaneous polarization effects.
	Besides, ZB QDs start emitting at lower energies than WZ QDs. 
\end{enumerate}

Further experiments are called for to confirm the above points, which should help improve
current understanding of the behavior and opportunities of these promising structures.

\appendix
\section{The effective Hamiltonian}\label{sec:effH}

The spin mixing observed in Section \ref{sec:hole} basically involves the energetically 
close HH up $|u_1\rangle$ and down $|u_4\rangle$ states. It originates from the 
spin-orbit interaction term $\Delta_{so}=\sqrt{2}\Delta_3$, as no mixing occurs if we set $\Delta_{so}=0$, 
due to the resulting block form of Hamiltonian (\ref{eq2}), 
thus preventing the interaction between spin up and spin down states. 
Also it is the result of a complex multiband interaction that cannot be reduced to 
an effective two band model. 
We can show it by reordering and splitting the basis vectors as follows, 
$\left\{\left\{|u_1\rangle, |u_4\rangle\right\},\left\{|u_2\rangle,|u_3\rangle,|u_5\rangle,|u_6\rangle\right\}\right\}$, 
that turns the Hamiltonian (\ref{eq2}) into:

\begin{equation}
\label{eq10}
 \left[ \begin{array}{cc|cccc} 
						F & 0 & -K^* & -H^* & 0 & 0    \\
						0 & F' & 0 &  0 &  -K & H  \\						
						\hline
						-K & 0 & G & H &  0 & \Delta_{so}  \\
						-H & 0 & H^* & \lambda & \Delta_{so} & 0\\
						0 & -K^* & 0 & \Delta_{so} & G' & -H^*   \\
						0 & H^* & \Delta_{so} & 0 & -H & \lambda' 
		 \end{array}  \right]  \equiv 
		  \left[ \begin{array}{c|c} 
		  H^{AA} & H^{AB} \cr
		  \hline
		  \cr
		  H^{BA} & H^{BB} \end{array}  \right]
\end{equation}
\noindent where $X=F,G,\lambda$ differs from $X'$ in a small $\kappa \mu_B B {\mathbb J}_z$ Zeeman term. %with $\kappa$ being the effective $g$ factor, $\mu_B$ the Bohr magneton, $B$ a small applied axial magnetic field and  ${\mathbb J}_z$  the angular momentum $z$-component diagonal matrix (with elements $\pm 3/2$, $\pm 1/2$).\\

\noindent The 2$\times$2 effective Hamiltonian $H^{eff}=H^{AA}+H^{AB}(I^{BB} \, E- H^{BB})^{-1} H^{BA}$, with $I^{BB}$ the $4 \times 4$ identity matrix, is usually approximated by setting $(H^{BB})_{ij}\approx E_i^{BB} \delta_{ij}$ that allows an straightforward  calculation of the inverse involved in the effective Hamiltonian, so that:
\begin{equation}
\label{eq11}
 (H^{eff})_{ij}= (H^{AA})_{ij} -\sum_{k \in B} \frac{H^{AB}_{ik} H^{BA}_{kj}}{E^{BB}_k-E}
\end{equation}

\noindent However, the particular form of $H^{AB}$ and $H^{BA}$ leads to zero extradiagonal elements for this approximate effective Hamiltonian. 

A more elaborate, but still simple, approximate effective Hamiltonian is obtained by setting $H\approx 0$,  $G\approx G'$ and  $\lambda \approx \lambda'$ in $H^{BB}$, thus yielding a twofold (cross-like) diagonal matrix. 
Its inverse $M=(I^{BB} \, E- H^{BB})^{-1}$ is still a twofold cross-like diagonal matrix and the product  $H^{AB} M H^{BA}$ is diagonal again. 

Similar results are obtained by employing the Lowdin perturbation theory to account for the action of  
$\left\{|u_2\rangle,|u_3\rangle,|u_5\rangle,|u_6\rangle\right\}$ on the Hamiltonian expanded in the $\left\{|u_1\rangle, |u_4\rangle\right\}$ basis set.\\

\noindent A multi-band Hamiltonian is needed to enable strong interaction between two states corresponding 
to the basis $i$ and $j$ despite $H_{ij}=0$. One of the simpler Hamiltonians illustrating this point would 
involve the basis set $\left\{|u_1\rangle,|u_2\rangle,|u_3\rangle\right\}$:
\begin{equation}
\label{eq12}
 		  \left[ \begin{array}{ccc} 
		  F & 0 & \Delta_1  \cr
		  0 & F' & \Delta_2 \cr
		  \Delta_1 & \delta_2 & X \end{array}  \right].
\end{equation}
\noindent $|u_1\rangle-|u_3\rangle$ mixing occurs at first order, $c_{13}^{(1)}=\frac{H_{31}}{\Delta E_{13}}$, while $|u_1\rangle-|u_2\rangle$ interaction holds at second order, $c_{12}^{(2)}=\frac{H_{23} H_{31}}{\Delta E_{12}\Delta E_{13}}=c_{13}^{(1)} \frac{H_{23}}{\Delta E_{12}} =\frac{\Delta_2}{F-F'}$. However, for a small Zeeman splitting $|F-F'| \ll \Delta_2$. 
Then, $c_{13}^{(1)} \ll c_{12}^{(2)}$.

%We note that smaller band-offsets have been proposed in other works,\cite{CorfdirNL}
%so ours is a lowerbound estimate of carrier tunneling into the barrier. % tmp ***

% Zhang&Zunger NL no considera Coulomb

%We stress again that in GaAs polytypes $V_{sp}$ is more important than $V_{pz}$,
%which is opposite to the case of many heterostructures like InGaN/GaN.\cite{ParkJAP}

\begin{acknowledgments}
We are grateful to P. Caroff, M.E. Pistol and B. Loitsch for useful discussions.
 Support from UJI project P1-1B2014-24, MINECO project CTQ2014-60178-P 
and a FPU grant (C.S.) is acknowledged.
\end{acknowledgments}


\begin{thebibliography}{30}
\bibitem{improve_mass} 
One could improve $H_h$, Eq.~(\ref{eqH1B}) by including the influence of decoupled bands on the HH effective mass. This can be done e.g. by fitting independently the WZ and ZB one-band mass parameters from a series of 6-band vs. one-band calculations. For the QDs we simulate, we have checked this leads to minor corrections. Since our study is of qualitative character, we have disregarded them.
\bibitem{why}
Using weakened $P_{sp}$ instead of null $P_{sp}$ allows us to calculate single particle
states confined in the wire phase, which form near the WZ/ZB interfase. These states would
otherwise spread all over the wire and form a poor starting point for the exciton self-consistent calculation.
\bibitem{self}
Upon inclusion of self-polarization potential, both curves in Fig.~\ref{fig7}(b)
would experience a similar blueshift due to the additional (dielectric) radial confinement of electron and hole.
Since the potential is mostly flat, the trends we describe would remain the same.
\bibitem{extra}
Unlike the calculations of Ref.~\onlinecite{VainoriusNL}, we consider Coulomb interaction and different masses for each crystal phase.
\bibitem{extra2}
Even for weakened $P_{sp}=2.3\cdot10^{-4}\,Cm^{-2}$, when the calculated redshift is similar to the experimental one,
we find a one-order-of-magnitude decrease in the electron-hole overlap.
\end{thebibliography}
\end{document}